
\input phyzzx
\twelvepoint
\font\small =cmr10 scaled 680
\font\bb=msbm10 scaled 1200
\font\bn=msam10 scaled 1200

\def\geq{\hbox{\bn\char '76}}
\def\leq{\hbox{\bn\char '66}}

\def\so{\hbox{\it\small 0}}

\def\bphi{{\bf\Phi}}
\def\bpsi{{\bf\Psi}}
\def\bfa{{\bf A}}
\def\bff{{\bf F}}

\def\tr{{\rm tr}}

\def\sun{$SU\!\left( N\right)$}
\def\sut{$SU\!\left( 2\right)$}

\def\picture #1 by #2 (#3){
	\vbox to #2{
	     \hrule width #1 height 0pt depth 0pt
	      \vfill
	      \special{picture #3}}}

\def\scaledpicture #1 by #2 (#3 scaled #4){{
	\dimen0=#1  \dimen1=#2
	\divide\dimen0 by 1000  \multiply\dimen0 by #4
	\divide\dimen1 by 1000  \multiply\dimen1 by #4
	\picture  \dimen0 by \dimen1 (#3 scaled #4)}}

\def\dalemb#1#2{{\vbox{\hrule height .#2pt
	\hbox{\vrule width.#2pt height#1pt \kern#1pt
		\vrule width.#2pt}
	\hrule height.#2pt}}}

\def\bogom{Bogomol'nyi}

\REF\vilenkin{T.W.B. Kibble, Journ. Phys. {\bf A9} (1976) 1387; A.
Vilenkin, Phys. Rev. {\bf D23} (1981) 852.}

\REF\anyon{Y.B. Zel'dovich, Mon. Not. R. Astron. Soc. {\bf 192}
(1980) 663; A. Vilenkin, Phys. Rev. Lett. {\bf 46} (1981) 1169; {\it
erratum} {\it ibid.} {\bf 46} (1981) 1496.}

\REF\paul{S.K. Paul and A. Khare, Phys. Lett. {\bf B174} (1986) 420;
C.N. Kumar and A. Khare, Phys. Lett. {\bf B178} (1986) 395; L. Jacobs,
A. Khare, C.N. Kumar and S.K. Paul, Int. Journ. Mod. Phys. {\bf A6}
(1991) 3441.}

\REF\hong{J. Hong, Y. Kim and P.Y. Pac, Phys. Rev. Lett. {\bf 64}
(1990) 2230; R. Jackiw and E.J. Weinberg, Phys. Rev. Lett. {\bf 64}
(1990) 2234.}

\REF\bogo{E.B. Bogomol'nyi, Sov. J. Nucl. Phys. {\bf 24} (1976) 449.}

\REF\jackiwt{R. Jackiw, K. Lee and E.J. Weinberg, Phys. Rev. {\bf
D42} (1990) 3488.}

\REF\nonstable{K. Lee, Phys. Rev. Lett. {\bf 66} (1991) 553; Phys.
Lett. {\bf B255} (1991) 381.}

\REF\flatt{H.J. de Vega and F.A. Schaposnik, Phys. Rev. Lett. {\bf
56} (1986) 2564.}

\REF\flatn{H.J. de Vega and F.A. Schaposnik, Phys. Rev. {\bf D34}
(1986) 3206.}

\REF\lozano{L.F. Cugliandolo, G. Lozano and F.A. Schaposnik, Phys.
Rev {\bf D40} (1989) 3440.}

\REF\cugl{L.F. Cugliandolo, G. Lozano, M.V. Man\'\i as and F.A.
Schaposnik, Mod. Phys. Lett. {\bf A6} (1991) 479.}

\REF\me{L.A.J. London, {\it Chern--Simons Vortices Coupled To
Gravity}, to appear in Physics Letters B.}

\REF\maximal{H.C. Tze and Z.F. Ezawa, Phys. Rev {\bf D14} (1976)
1006; H.J. de Vega, Phys. Rev. {\bf D18} (1978) 2932; P. Hasenfratz,
Phys. Lett. {\bf B85} (1979) 338; A.S. Schwarz and Y.S. Tyupkin, Phys.
Lett. {\bf B90} (1980) 135.}

\REF\deser{S. Deser, R. Jackiw and S. Templeton, Phys. Rev. Lett.
{\bf 48} (1982) 975; Ann. Phys. {\bf 140} (1982) 370.}

\REF\linet{B. Linet, Gen. Rel. Grav. {\bf 22} (1990) 469.}

\date={}
\Pubnum={GCR--95/06/02}

\titlepage

\title{Non--Abelian, Self--Dual Chern--Simons Vortices Coupled To
Gravity}

\centerline{M.E.X. Guimar\~aes\foot{marg@ccr.jussieu.fr} and L.A.J.
London\foot{london@ccr.jussieu.fr}}

\address{Laboratoire de Gravitation et Cosmologie Relativistes\break
	 Universit\'e Pierre et Marie Curie, CNRS/URA 769\break
	 Tour 22/12, ${4}^{\rm\scriptscriptstyle \grave eme}$
	 \'etage, Bo\^\i te 142\break
	 4, Place Jussieu\break
	 75252, Paris Cedex 05\break
	 France}

\vfil

\abstract

In this article we consider \sut\ Chern--Simons/Higgs theory coupled
to gravity in three dimensions. It is shown that for a cylindrically
symmetric vortex both the Einstein equations and the field equations
can be reduced to a set of first--order \bogom\ equations provided,
that we choose a specific eighth--order potential.

\endpage

\chapter{Introduction}

Vortices, monopoles and other topological defects are regular,
classical solutions to gauge field theories which arise when a symmetry
of the theory is spontaneously broken. In particular, vortices may be
regarded as cosmic strings generated during phase transitions in the
early universe [\vilenkin] and they can provide the seeds which are
required for the formation of galaxies [\anyon]. Vortices arise
whenever a gauge group, $G$ is spontaneously broken to a disconnected
unbroken subgroup, $H$. The simplest gauge theory that admits cosmic
string (vortex) solutions is Abelian Higgs theory. Here a complex
Higgs scalar field, $\phi$ self--interacts via a fourth--order gauge
invariant Higgs potential. After the $U\!\left( 1\right)$ symmetry is
spontaneously broken, the vacuum is invariant under the action of $I$
(which is a subgroup of $U\!\left( 1\right)$), and it is characterised
by a non--vanishing expectation value for the Higgs field.
Mathematically, this situation corresponds to a topologically
non--trivial vacuum with first homotopy group, $\pi_1\left(U\!\left(
1\right) /I\right)=\hbox{\bb Z}$.

Recently there has been a considerable amount of interest in
Maxwell/Chern--Simons /Higgs theory in three dimensional Minkowski
space [\paul], due to its similarities with the theory of high--$T_c$
superconductors. It was first remarked in [\hong] that at large
distances the Chern--Simons term dominates the Maxwell term and so it
is reasonable to consider the simpler Abelian Chern--Simons/Higgs
theory. Consequently, it was shown in [\hong] that there exist vortex
solutions to three dimensional Abelian Chern--Simons/Higgs theory and
moreover, by choosing a specific sixth--order potential the field
equations reduce to a set of first--order self--dual (Bogomol'nyi
[\bogo]) equations. This potential has a symmetric and an
antisymmetric vacuum. Solutions approaching the antisymmetric vacuum
at infinity describe topologically stable vortices, whereas solutions
which approach the symmetric phase at infinity are non--topological
solutions [\jackiwt].

The most natural generalisation of the Abelian theories above is to
examine their non--Abelian counterparts. In [\nonstable] it was
demonstrated that there exist self--dual vortex solutions to \sun\
Chern--Simons/Higgs theory in three dimensional Minkowski space. As
is to be expected from our knowledge of the Abelian case, a
sixth--order potential is required in order to obtain a set of
\bogom\ equations, but in contrast to the Abelian case there are no
topologically stable solutions, only stable non--topological
solutions in both the symmetric and antisymmetric phases. In [\flatt]
and [\flatn], flat space Yang--Mills/Chern--Simons/Higgs theories were
investigated for the gauge groups \sut\ and \sun\ respectively. By
including several Higgs multiplets in the theory (to ensure the
maximal breaking of the gauge symmetry) topologically stable vortex
solutions were found. However, since the models considered contained
only a fourth--order potential these solutions were not shown to be
self--dual. Self--dual vortex solutions to \sun\
Yang--Mills/Higgs theory with $N$ Higgs multiplets and a sixth--order
potential are found in [\lozano]. An interesting feature of these
self--dual vortices is that the \bogom\ bound upon the mass of the
configuration in terms of its topological charge is {\it not} a
topologically invariant quantity, in contrast to the Abelian case.
The difference in the values of the \bogom\ bound associated with
members of the same homotopy class is related to the fact that the
topological charge of \sun\ vortices is defined modulo $N$, whereas
physical quantities may depend upon the actual value of the magnetic
flux associated with this topological charge [\lozano]. In [\cugl],
\sut\ Chern--Simons/Higgs theory was examined and, by using a
particular choice of sixth--order potential, self--dual vortex
solutions were obtained. As in [\lozano], these vortices  obey a
\bogom\ bound which is not a topologically invariant quantity.

Another interesting generalisation of [\paul , \hong] is to couple the
vortex solutions to gravity. In [\me], Abelian Chern--Simons/Higgs
theory was coupled to three dimensional gravity. It was demonstrated
that there exist vortex solutions such that, by choosing a
non--renormalisable eighth--order potential whose constant parameters
were precisely chosen, {\it both} the Einstein equations and the
equations of motion can be reduced to a set of first order \bogom\
equations. In this article we will search for self--dual vortex
solutions to \sut\ Chern--Simons/Higgs theory coupled to three
dimensional gravity. We will show that both the mass and the angular
momentum of configurations belonging to the same homotopy class can
take different values due to the mathematical considerations explained
below. This means that neither the mass nor the angular momentum are
topologically invariant.

This article is organised as follows: in Section $2$ we describe the
mathematical basis of the maximal breaking of an \sun\ gauge symmetry,
as well as describing  several unusual features of non--Abelian gauge
theories. In Section $3$ we present a review of Einstein's theory of
gravity in three dimensions which will serve to define the notation
that will be used throughout this article. In Section $4$ we derive a
set of first--order Bogomol'nyi equations from the Einstein equations
and the equations of motion of three dimensional
Einstein/Chern--Simons/Higgs theory. In Section $5$ we present our
conclusions.

\chapter{Charged Vortices In \sun\ Gauge Theories}

In this article we will consider classical gauge theories with gauge
group $G$ (this will subsequently chosen to be either \sut\ or \sun
). Upon spontaneous symmetry breaking the symmetry group is reduced
to an unbroken group $H$ which is a subgroup of $G$. This spontaneous
symmetry breaking is achieved via the Higgs mechanism using a
symmetry breaking potential, $V$ whose zeros can be identified with
the coset space $G/H$. In $\left( d+1\right)$--dimensions, in order
to have topologically stable solutions the $\left( d-1\right)^{\rm
th}$ homotopy group of $G/H$, $\pi_{d-1}\left( G/H\right)$ must be
non--trivial, \ie\ it must have more than one element. This is
because the connected components of the space of non--singular,
finite energy solutions are in $1$--to--$1$ correspondence with the
homotopy classes of mappings from the $\left( d-1\right)$--sphere,
${\cal S}^{d-1}$ to $G/H$.

For example, in the Abelian case $G=U\!\left( 1\right)$ and $H=I$ (in
fact the vacuum manifold is topologically equivalent to ${\cal S}^1$).
Thus $\pi_1\left( G/H\right)\equiv\pi_1\left( U\!\left( 1\right)\right)
=\hbox{\bb Z}$ and so there exists an infinity of topologically stable
vortices labelled by an integer, $n$ ($n=0$ labels the vacuum). We are
interested in the non--Abelian case where $G=\,$\sun . We consider
theories for which the Higgs fields are in the adjoint representation
of \sun\ and we will assume maximal symmetry breaking of \sun . This
means that the vacuum is only invariant under the unit matrix in the
adjoint representation. There exist $N$ elements with this property,
namely the matrices at the centre of \sun
$$
I_{\!\scriptscriptstyle N}e^{{2\pi in\over N}}\ ,
\qquad
n =0,1,\dots ,N-1\ ,
\eqn\ceneqn
$$
where $I_{\!\scriptscriptstyle N}$ is the $N\times N$ unit matrix.
Hence, since the roots of unity $e^{{2\pi in\over N}}$ provide a
representation of the Abelian group, $\hbox{\bb
Z}_{\!\scriptscriptstyle N}$ we see that $H=\hbox{\bb
Z}_{\!\scriptscriptstyle N}$. Thus
$$
\pi_1\left( G/H\right)
\equiv
\pi_1\left(\hbox{\sun}/\hbox{\bb Z}_{\!\scriptscriptstyle N}\right)
=
\hbox{\bb Z}_{\!\scriptscriptstyle N}\ .
\eqn\homoeqn
$$
This implies that there exist $\left( N-1\right)$ topologically stable
vortex solutions (recall that $N=0$ labels the vacuum solution). An
element of each of the non--trivial classes above can be obtained from
the vacuum class by a non--trivial gauge rotation, $g_n\in\,$\sun.
Since we are concerned with three dimensional theories, the direction
at infinity is characterised by an angle, $\theta$. A map belonging to
the $n^{\rm th}$ homotopy class satisfies, upon performing one turn
around a closed contour
$$
g_n\left( 2\pi\right)=e^{{2\pi in\over N}}g_n\left( 0\right)\ ,
\qquad
n =0,1,\dots ,N-1\ ,
\eqn\gnvalue
$$
where $g_n$ is an element of the Cartan subgroup of \sun . A peculiar
feature of these theories with maximal symmetry breaking is that, in
order to obtain topologically stable vortex solutions, the theories
must contain $N$ Higgs multiplets [\flatt , \flatn , \maximal].

Another unusual property of non--Abelian Chern--Simons theories is
that if one wishes to have a consistent path integral formulation of
the theory, the Chern--Simons coupling constant, $\kappa$ is quantised.
Under `small' gauge transformations, \ie\ gauge transformations which
are connected to the identity, the Chern--Simons,
$S_{\!\scriptscriptstyle CS}$ is gauge invariant. However, under
`large' gauge transformations (those gauge transformations that are not
connected to the identity) $S_{\!\scriptscriptstyle CS}$ is not gauge
invariant. In fact
$$
S_{\!\scriptscriptstyle CS}\rightarrow S_{\!\scriptscriptstyle CS} -
{8\pi^2\kappa\over e^2}\omega\left( U\right)\ ,
\eqn\noninv
$$
where $e$ is the gauge field coupling constant and $\omega\left(
U\right)$ is the integer--valued winding number of the gauge
transformation, $U\in\,$\sun. Explicitly, $\omega\left( U\right)$ is
given by
$$
\omega\left( U\right)={1\over 24\pi^2}\int\!
d^3x\varepsilon^{\mu\nu\lambda}\tr\left(
U^{-1}\partial_{\mu}UU^{-1}\partial_{\nu}UU^{-1}\partial_{\lambda}U
\right)\ .
\eqn\winding
$$
Since, for a consistent path integral formulation ${\rm exp}\left(
iS_{\!\scriptscriptstyle CS}\right)$ is required to be gauge
invariant we see that
$$
-{8\pi^2\kappa\over e^2}=2\pi m\ ,
\qquad
m\in\hbox{\bb Z}\ ,
\eqn\quan
$$
which implies that $\kappa$ is quantised in units of $-{e^2\over
4\pi}$ [\deser], \ie\
$$
\kappa =-{me^2\over 4\pi}\ .
\eqn\kappaquan
$$

\chapter{Einstein's Theory Of Gravity In Three Dimensions}

Einstein's theory of gravity in three dimensions exhibits some rather
interesting behaviour which makes its analysis as important as that
of the more usual four dimensional case. In the absence of any matter
sources it turns out that, in a topologically trivial three
dimensional spacetime, the theory is trivial. However, upon the
introduction of a point (or line) matter source the spacetime
acquires a global conical structure and non--trivial gravitational
effects can occur in the framework of both classical and quantum
theories.

Throughout this article we will use the notation adopted in [\me,
\linet]. Let us consider a three dimensional spacetime with metric
$g_{\mu\nu}$ and coordinates $x^{\mu}=\left(\rho ,t,\theta\right)$
where $t$ is the timelike coordinate. The metric has signature $\left(
-,+,-\right)$. A stationary, cylindrically symmetric three dimensional
spacetime can be described by the line element
$$
ds^2=-d\rho^2+g_{ij}\left(\rho\right)
dx^idx^j\ , \qquad\rho\geq 0\ ,
\qquad 0\leq\theta <2\pi\ ,
\qquad i,j=\left( t,\theta\right)\ ,
\eqn\linelm
$$
In what follows we will choose the coordinates such that $x^t=t$,
$x^\theta=\theta$. Note that the inverse metric, $g^{\mu\nu}$ is given
by
$$
g^{\mu\nu}={1\over g}\pmatrix{-g&0&0\cr
0&g_{\theta\theta}&-g_{t\theta}\cr
0&-g_{t\theta}&g_{tt}\cr}\ ,
\eqn\inversemet
$$
where the determinant of the metric is given by
$$
\det g_{\mu\nu}
=g_{t\theta}^{2}-g_{tt}g_{\theta\theta}=-g\equiv -\det g_{ij}\ .
\eqn\detmet
$$
Note that in order to obtain a metric of the required
(pseudo--Riemannian) signature we impose the conditions $g_{tt}>0$ and
$g_{\theta\theta}<0$. As was explained in [\me, \linet], the metric is
required to have the following asymptotic behaviour
$$
\eqalign{
ds^2&\sim -d\rho^2+dt^2-2\omega\rho^2dtd\theta-\rho^2d\theta^2
\qquad{\rm as}\ \rho\rightarrow 0
\cr
ds^2&\sim -d\rho^2+\left(Adt+{4GJ\over
B}d\theta\right)^2-B^2\rho^2d\theta^2
\qquad{\rm as}\ \rho\rightarrow\infty\ ,}
\eqn\asymbehav
$$
where $A$, $B$, $G$, $J$ and $\omega$ are constants\foot{Note that in
three dimensions there is a certain ambiguity over the sign of
Newton's constant, $G$. For simplicity we will assume that $G$ is
positive.}. The second of the metrics \asymbehav\ (as a vacuum metric)
describes a particle of mass $M$ and spin $J$ located at $\rho =0$,
where $M$ is given by $B=1-4GM$. Furthermore, note that for this metric
we have
$$
g=-\left( AB\rho\right)^2\ .
\eqn\detvalue
$$

We now define the quantities
$$
\chi^i{}_j\equiv g^{ik}{d\over d\rho}g_{kj}\ ,
\eqn\chidef
$$
such that
$$
\tr\chi^i{}_j\equiv\chi^i{}_i={2\over\sqrt{-g}}{d\over d\rho}\sqrt{-g}\
{}.
\eqn\trdef
$$
It is well known that in three dimensions the Einstein equations
$\left( G_{\mu\nu}=8\pi GT_{\mu\nu}\right)$ can be expressed in terms
of the $\chi^i{}_j$ as
$$
{1\over\sqrt{-g}}{d\over d\rho}\left(\sqrt{-g}\chi^i{}_j\right)=
16\pi G\left( T^i{}_j-T\delta^i{}_j\right)\ ,
\eqn\exeinstein
$$
and
$$
G^{\rho}{}_{\rho}=-{1\over 4}\det\chi^i{}_j=8\pi GT^{\rho}{}_{\rho}\ .
\eqn\rhocpt
$$

Using the form of the metric near $\rho =0$ given by the first of
equations \asymbehav\ we obtain the relations (for regular matter
fields)
$$
\sqrt{-g}\chi^t{}_{\theta}=16\pi
G\int^{\rho}_{\so}\!\sqrt{-g}T^t{}_{\theta}d\rho\ ,
\qquad
\sqrt{-g}\chi^{\theta}{}_t=2\omega +16\pi
G\int^{\rho}_{\so}\!\sqrt{-g}T^{\theta}{}_t d\rho\ .
\eqn\zeroeqn
$$
Using the second of the metrics \asymbehav\ we find that as $\rho$
tends to infinity the quantities on the right--hand side of equations
\zeroeqn\ are given by
$$
\sqrt{-g}\chi^t{}_{\theta}=-8GJ\ ,
\qquad
\sqrt{-g}\chi^{\theta}{}_t=0\ ,
\eqn\infeqn
$$
and hence we obtain the following expressions for $J$ and $\omega$
$$
J=-2\pi\int^{\infty}_{\so}\!\sqrt{-g}T^t{}_{\theta}d\rho\ ,
\qquad
\omega=-8\pi G\int^{\infty}_{\so}\!\sqrt{-g}T^t{}_{\theta}d\rho\ .
\eqn\omegjeqn
$$

\chapter{\sut\ Chern--Simons/Higgs Vortices Coupled To Gravity}

In this section we search for vortex solutions to \sut\
Chern--Simons/Higgs theory coupled to gravity. To ensure maximal
symmetry breaking we consider a theory containing two Higgs
multiplets, $\Phi$ and $\Psi$ with vacuum expectation values $\eta$
and $\xi$ respectively. We see from equation \homoeqn\ that here
(since $G=\,$\sut ) $\pi_1\left( G/H\right) =\hbox{\bb
Z}_{\!\scriptscriptstyle 2}$ and so there exists only one class of
topologically stable vortex solutions. Furthermore, an element $g$ of
the Cartan subgroup of \sut\ can be represented by
$$
g=e^{i\sigma_3\Omega\left(\theta\right)}\ ,
\eqn\grepn
$$
where $\sigma_3$ is the third Pauli matrix and
$\Omega\left(\theta\right)$ obeys
$$
\Omega\left( 2\pi\right) -\Omega\left( 0\right)
=\left( n+2s\right)\pi\ ,
\qquad
n=0,1\ ,
\qquad
s\in\hbox{\bb Z}\ .
\eqn\omegaval
$$

In order to obtain solutions with finite energy we require that the
gauge and Higgs fields have the following asymptotic behaviour
$$
\eqalign{
\lim_{\rho\rightarrow\infty}D_{\mu}\Phi &=
\lim_{\rho\rightarrow\infty}D_{\mu}\Psi =0
\cr
\lim_{\rho\rightarrow\infty}A_{\mu}&=\lim_{\rho\rightarrow\infty}
A_{\mu}^a\sigma_a ={1\over i}g^{-1}\partial_{\mu}g\ ,}
\eqn\finite
$$
where $a$ is an \sut\ index and the $\sigma_a$ are the Pauli
matrices. Note that the second of these conditions implies that
$A_{\mu}$ is pure gauge at infinity and so, at infinity, $F_{\mu\nu}$
vanishes. Upon combining the second of the above conditions with the
form of $g$ given by equation \grepn\ we find that
$$
\lim_{\rho\rightarrow\infty}A_{\theta}^3={d\Omega\left(\theta\right)
\over d\theta}\ ,
\eqn\limtheta
$$
whilst all the other components of $A^a_{\mu}$ vanish at infinity.

The action describing this theory is given by
$$
\eqalign{
S&=\int\! d^3x\sqrt{-g}\Biggl(
-{1\over 16\pi G}R
+{1\over
2}\left[D_{\mu}{\bf \Phi}\cdotp D_{\nu}{\bf \Phi}
g^{\mu\nu}\right] +{1\over 2}\left[ D_{\mu}{\bf\Psi}\cdotp
D_{\nu}{\bf\Psi} g^{\mu\nu}\right]
\cr
&+{\kappa\over 4\sqrt{-g}}\varepsilon^{\mu\nu\lambda}
\left[
{\bf F}_{\mu\nu}\cdotp{\bf A}_{\lambda}
-{2e\over 3}{\bf A}_{\mu}\cdotp\left({\bf A}_{\nu}\times{\bf
A}_{\lambda}\right)\right] -V\!\left({\bf\Phi},{\bf\Psi}\right)
\Biggr)}
\eqn\gsutact
$$
where\foot{Note
that $R=g^{\mu\nu}R_{\mu\nu}=g^{\mu\nu}R^{\lambda}{}_{\mu\lambda\nu}$
and $\varepsilon^{\mu\nu\lambda}$ is the totally antisymmetric
Levi--Civita tensor density.} the \sut\ covariant derivative and field
strength are respectively given by
$$
\eqalign{
D_{\mu}&=\partial_{\mu}+e\bfa_{\mu}\times
\cr
\bff_{\mu\nu}&=\partial_{\mu}\bfa_{\nu}-\partial_{\nu}\bfa_{\mu}+
e\bfa_{\mu}\times\bfa_{\nu}\ .}
\eqn\dervdef
$$
Note that the vector notation in the above quantities refers to the
internal \sut\ vectors. The Einstein equations of the theory are
$$
G_{\mu\nu}=8\pi T_{\mu\nu}\ ,
\eqn\eineqnsut
$$
where
$$
\eqalign{
T_{\mu\nu}&=\left( D_{\mu}\bphi\cdotp D_{\nu}\bphi -{1\over
2}g_{\mu\nu}D_{\alpha}\bphi\cdotp D_{\beta}\bphi g^{\alpha\beta}\right)
\cr
&+
\left( D_{\mu}\bpsi\cdotp D_{\nu}\bpsi -{1\over
2}g_{\mu\nu}D_{\alpha}\bpsi\cdotp D_{\beta}\bpsi
g^{\alpha\beta}\right) +g_{\mu\nu}V\!\left(\bphi ,\bpsi\right)\ .}
\eqn\emsut
$$
The field equations associated with the variations of $\bphi$,
$\bpsi$ and $\bfa_{\mu}$ are given by
$$
\eqalign{
D_{\mu}\left(\sqrt{-g}g^{\mu\nu}D_{\nu}\bphi\right)&=
-\sqrt{-g}{\partial V\over\partial\bphi}
\cr
D_{\mu}\left(\sqrt{-g}g^{\mu\nu}D_{\nu}\bpsi\right)&=
-\sqrt{-g}{\partial V\over\partial\bpsi}
\cr
{\kappa\over
2}\varepsilon^{\mu\nu\lambda}\bff_{\nu\lambda}&=
-e\sqrt{-g}g^{\mu\delta}
\left(
D_{\delta}\bphi\times\bphi +D_{\delta}\bpsi\times\bpsi
\right)\ .}
\eqn\fieldsut
$$

Guided by work on self--dual $U\left(\! 1\right)$ vortices coupled to
gravity [\me] and flat space self--dual \sut\ vortex solutions
[\cugl], we search for solutions to the Einstein equations and
the field equations with a potential of the form
$$
\eqalign{
V\!\left({\bf\Phi},{\bf\Psi}\right)
&=
\alpha{\bf\Phi}^2\left({\bf\Phi}^2-\eta^2\right)^2
+
\beta\left({\bf\Phi}^2-\eta^2\right)^4
\cr
&+
\gamma{\bf\Psi}^2\left({\bf\Psi}^2-\xi^2\right)^2
+
\delta\left({\bf\Psi}^2-\xi^2\right)^4
\cr
&+
\lambda\left({\bf \Phi}\cdotp{\bf\Psi}\right)^2\ ,}
\eqn\veqn
$$
where $\alpha$, $\beta$, $\gamma$, $\delta$ and $\lambda$ are
constants to be determined\foot{Note that we can of course write a
more general eighth--order potential but upon imposing the conditions
described by equations $\left( 4.11\right)$ we see that these terms
vanish. Hence, for clarity we include only the first few possible
terms.}. It has been shown in flat space [\flatt , \cugl] that \sut\
vortices can be obtained if one assumes that the only r\^ ole played
by one of the Higgs fields is to ensure maximal symmetry breaking,
\ie\ the Higgs field is taken to be constant over all of the spacetime
and hence has no dynamical r\^ ole. Since we are searching for vortex
solutions that correspond to a set of \bogom\ equations (\ie\ a
minimal energy configuration) it is reasonable to assume (as in flat
space [\flatt , \cugl]) that any configuration with non--constant
$\bpsi$ has greater energy than a configuration with constant $\bpsi$.
Hence, we impose the following conditions (valid over all of the
spacetime) upon $\bpsi$
$$
\eqalign{
\bpsi^2&=\xi^2
\cr
D_{\mu}\bpsi&=0
\cr
\bphi\cdotp\bpsi&=0\ .}
\eqn\psicond
$$
The last of these conditions ensures that $\bphi$ and $\bpsi$ are
not parallel in the internal space. Upon imposing these conditions we
find that the second of the field equations \fieldsut\ is
automatically satisfied, whereas the remaining equations reduce to
$$
\eqalign{
D_{\mu}\left(\sqrt{-g}g^{\mu\nu}D_{\nu}\bphi\right)&=
-\sqrt{-g}{\partial V\over\partial\bphi}
\cr
{\kappa\over
2}\varepsilon^{\mu\nu\lambda}\bff_{\nu\lambda}&=
-e\sqrt{-g}g^{\mu\delta}
\left(
D_{\delta}\bphi\times\bphi\right)\ .}
\eqn\newfield
$$
Furthermore, the energy--momentum tensor and the potential are given by
$$
T_{\mu\nu}=\left( D_{\mu}\bphi\cdotp D_{\nu}\bphi -{1\over
2}g_{\mu\nu}D_{\alpha}\bphi\cdotp D_{\beta}\bphi
g^{\alpha\beta}\right) +g_{\mu\nu}V\!\left(\bphi\right)\ ,
\eqn\newem
$$
and
$$
V\!\left(\bphi\right)
=
\alpha\bphi^2\left(\bphi^2-\eta^2\right)^2
+
\beta\left(\bphi^2-\eta^2\right)^4\ ,
\eqn\newpot
$$
respectively.

As we remarked above, for gauge group \sut\ there is only one class
of topologically stable vortex solutions ($n=1$). Thus, we search for a
cylindrically symmetric vortex solution of the form
$$
\eqalign{
\bfa_{\rho}&={\bf 0}\ ,
\qquad
\bfa_t=\epsilon{W\left(\rho\right)\over e}\pmatrix{0\cr 0\cr 1\cr}\ ,
\qquad
\bfa_{\theta}={\left( P\left(\rho\right) -n-2s\right)\over
e}\pmatrix{0\cr 0\cr 1\cr}
\cr
\bphi&=R\left(\rho\right)\pmatrix{\cos\theta\cr\sin\theta\cr 0\cr}\ ,
\qquad
\bpsi =\xi\pmatrix{0\cr 0\cr 1\cr}\ ,}
\eqn\gaugeans
$$
where $\epsilon =\pm 1$\foot{The sign of $\epsilon$ will be chosen
depending upon the sign of $\left( n+2s\right)$ to obtain a positive
mass.}. Note that the presence of the integer $s$ in the expression
for $\bfa_{\theta}$ is necessary for the {\it ansatz} to be
consistent with equation \limtheta . We also require that the fields
obey the boundary conditions at the origin
$$
P\left( 0\right) =n\ ,
\qquad
R\left( 0\right) =0\ ,
\eqn\orgbdy
$$
which ensures that the fields are single valued. The finite energy
boundary conditions given by equations \finite\ imply that we must
impose the following boundary conditions at infinity
$$
P\left(\infty\right) =0\ ,
\qquad
R\left(\infty\right) =\eta\ ,
\qquad
W\left(\infty\right) =0\ .
\eqn\inftbdy
$$
Furthermore, to enable us to obtain a solution we henceforth assume
that $g_{tt}=1$ and thus in equation \asymbehav , $A=1$.

Substituting the {\it ans\"atze} \gaugeans\ into the field equations
\newfield\ we obtain
$$
\eqalign{
\pmatrix{\cos\theta\cr\sin\theta\cr 0\cr}\left[
{1\over\sqrt{-g}}{d\over d\rho}\left(\sqrt{-g}{dR\over d\rho}\right)
+
{R\over g}\left( P^2+W^2g_{\theta\theta}\right)
-
{2\epsilon\over g}PRWg_{t\theta}\right]
&={\partial V\!\left(\bphi\right)\over\partial\bphi}
\cr
\pmatrix{0\cr 0\cr 1\cr}\left[
{\kappa\over\sqrt{-g}}{dP\over d\rho}
+
{e^2R^2\over g}\left(\epsilon
Wg_{\theta\theta}-Pg_{t\theta}\right)\right]&=0
\cr
\pmatrix{0\cr 0\cr 1\cr}\left[
{\epsilon\kappa\over\sqrt{-g}}{dW\over d\rho}
-
{e^2R^2\over g}\left( P-\epsilon Wg_{t\theta}\right)\right]&=0\ .}
\eqn\redfield
$$
Furthermore, it is easy to see from \newpot\ that
$$
\eqalign{
{\partial V\!\left(\bphi\right)\over\partial\bphi}&=
2\pmatrix{\cos\theta\cr\sin\theta\cr 0\cr}R\left( R^2-\eta^2\right)
\left[\alpha\left( R^2-\eta^2\right) +2\alpha R^2 +4\beta
\left( R^2-\eta^2\right)^2\right]
\cr
&\equiv\pmatrix{\cos\theta\cr\sin\theta\cr 0\cr}H\left( R\right)
\ ,}
\eqn\dveqn
$$
and thus, since the field equations are valid for all $\theta$, the
field equations \redfield\ become
$$
\eqalign{
{1\over\sqrt{-g}}{d\over d\rho}\left(\sqrt{-g}{dR\over d\rho}\right)
+
{R\over g}\left( P^2+W^2g_{\theta\theta}\right)
-
{2\epsilon\over g}PRWg_{t\theta}
&=
H\left( R\right)
\cr
{\kappa\over\sqrt{-g}}{dP\over d\rho}
+
{e^2R^2\over g}\left(\epsilon Wg_{\theta\theta}-Pg_{t\theta}\right)&=0
\cr
{\epsilon\kappa\over\sqrt{-g}}{dW\over d\rho}
-
{e^2R^2\over g}\left( P-\epsilon Wg_{t\theta}\right)&=0\ .}
\eqn\fredfield
$$
We now notice the these field equations are identical to the field
equations $(3.9)$ of [\me] for Abelian Chern--Simons vortices coupled
to gravity. Hence, proceeding as in [\me] we assume that
$W\left(\rho\right)$ is given by
$$
W\left(\rho\right) ={e^2\over 2\kappa}\left( R^2-\eta^2\right)\ .
\eqn\wform
$$
Thus, the third of the field equations \fredfield\ implies that
$$
\sqrt{-g}{dR\over d\rho}=RWg_{t\theta}-\epsilon RP\ .
\eqn\drform
$$
In order to satisfy the first of the above field equations we find
that (using the last two field equations)
$$
\alpha ={e^4\over 8\kappa^2}\ ,
\qquad
{1\over\sqrt{-g}}{dg_{t\theta}\over d\rho}={16\beta\kappa\over e^2}
\left( R^2-\eta^2\right)^2\ .
\eqn\alphaform
$$

To pursue this analysis any further, which will enable us to determine
the values of $\beta$ and ${dg_{\theta\theta}\over d\rho}$, we must
examine the Einstein equations \eineqnsut . Using equation \newem\ for
the energy--momentum tensor we calculate $T^{\theta}{}_t$ and
$T^t{}_{\theta}$. We simplify the resulting expressions using the third
of the field equations \fredfield\ to obtain
$$
T^{\theta}{}_t={\kappa\over e^2\sqrt{-g}}W{dW\over d\rho}\ ,
\qquad
T^t{}_{\theta}=-{\kappa\over e^2\sqrt{-g}}P{dP\over d\rho}\ .
\eqn\tforms
$$
Substituting these two equations into equation \exeinstein\ and
integrating yields respectively
$$
\sqrt{-g}\chi^{\theta}{}_t={8\pi G\kappa\over e^2}W^2+C\ ,
\qquad
\sqrt{-g}\chi^{t}{}_{\theta}=-{8\pi G\kappa\over e^2}P^2+D\ ,
\eqn\chivals
$$
where $C$ and $D$ are constants of integration. We now examine the
equation for $\chi^{\theta}{}_t$ as $\rho\rightarrow\infty$. Using
the metric and boundary conditions near infinity we find that
$$
\sqrt{-g}\chi^{\theta}{}_t={8\pi G\kappa\over e^2}W^2\ .
\eqn\nchival
$$
Moreover, investigating the equation for $\chi^{t}{}_{\theta}$
near the origin yields
$$
\sqrt{-g}\chi^{t}{}_{\theta}=-{8\pi G\kappa\over e^2}\left(
P^2-n^2\right)\ .
\eqn\fcval
$$
Hence, using the expressions for $J$ and $\omega$ given by equations
\omegjeqn\ we obtain
$$
J=-{\pi\kappa\left( n+2s\right)^2\over e^2}\ ,
\qquad
\omega ={\pi Ge^2\over\kappa}\eta^4\ .
\eqn\jval
$$
In general, $\chi^{\theta}{}_t=g^{-1}{dg_{t\theta}\over d\rho}$. Thus
combining equations \alphaform\ and \nchival\ yields
$$
\beta =-{\pi Ge^4\over 8\kappa^2}\ ,
\eqn\betval
$$
and so the potential is given by
$$
V\!\left(\bphi\right)
={e^4\over 8\kappa^2}\left(\bphi^2-\eta^2\right)^2
\left[\bphi^2-\pi G\left(\bphi^2-\eta^2\right)^2\right]\ .
\eqn\sdpot
$$

Without this specific form for $V\!\left(\bphi\right)$ one cannot
obtain a set of Bogomol'nyi equations, which justifies the {\it
ansatz} \veqn\ for the potential. We note that, as anticipated in
the introduction, this eighth--order potential is non--renormalisable
and furthermore, the self--dual solutions with
$\lim_{\rho\rightarrow\infty}\Phi^2=\eta^2$ are locally stable and
the potential is unbounded below as $\Phi\rightarrow\infty$. As in the
Abelian case discussed in [\me], one of the principal differences
between this theory and the flat space theory discussed in [\cugl] is
that here the required potential is of eighth--order, whereas in flat
space a sixth--order potential was required to obtain a set of
Bogomol'nyi equations. Another distinguishing feature of the model
considered here is the lack of non--topological solutions. In flat
space it was demonstrated in [\jackiwt, \nonstable] that there exist
solutions that asymptotically approach the symmetric vacuum of the
potential. These solutions are known as non--topological solutions
since the topology of the symmetric vacuum is trivial. For the model
under consideration here there do not exist any non--topological
solutions since there no longer exists a symmetric vacuum, due to the
presence of the eighth--order term in the potential. Note that there
exists the possibility of a solution corresponding to the minimum of
the potential at $\Phi^2 =0$, but this is not of the type of solutions
considered here.

Note that (using equation \emsut )
$$
\eqalign{
T^t{}_t-T^{\theta}{}_{\theta}&
=
{R^2\over g}\left( W^2g_{\theta\theta}-P^2\right)
\cr
&=
-{\epsilon\kappa\over e^2\sqrt{-g}}\left( W{dP\over d\rho}+P{dW\over
d\rho}\right)\ ,}
\eqn\twots
$$
where we have used equations \fredfield\ to obtain the second equality.
Therefore, equation \exeinstein\ yields (upon integrating)
$$
{1\over\sqrt{-g}}{dg_{\theta\theta}\over d\rho}=
-{16\pi G\epsilon\kappa\over e^2}PW+E\ ,
\qquad
E\in\hbox{\bb R}\ .
\eqn\geqn
$$
Near $\rho =0$, $g_{\theta\theta}\sim -\rho^2$. Consequently, it is
straightforward to see that
$$
{1\over\sqrt{-g}}{dg_{\theta\theta}\over d\rho}=
-\epsilon 8\pi G\left[
P\left( R^2-\eta^2\right) +\left( n+2s\right)\eta^2\right] -2\ .
\eqn\fgeqn
$$
As $\rho\rightarrow\infty$, $g_{\theta\theta}\sim -B^2\rho^2$. Thus,
using equation \fgeqn\ (as $\rho\rightarrow\infty$) we obtain
$$
B=1+\epsilon 4\pi G\left( n+2s\right)\eta^2\ .
\eqn\bval
$$
Since, as was stated in Section $3$, $B=1-4GM$, we find that
$$
M=-\epsilon\pi\left( n+2s\right)\eta^2\ .
\eqn\mval
$$

There remains one component of Einstein's equation to examine, namely
equation \rhocpt . Using equations \wform , \drform\ and \sdpot\ we see
that $$
T^{\rho}{}_{\rho}=-{2\pi G\kappa^2\over e^4}W^4\ .
\eqn\trr
$$
Furthermore,
$$
\eqalign{
\det\chi^i{}_j&=\chi^{\theta}{}_{\theta}\chi^t{}_t-
\chi^t{}_{\theta}\chi^{\theta}{}_t
\cr
&=-{1\over g}\left({dg_{t\theta}\over d\rho}\right)^2\ ,}
\eqn\deteqn
$$
where we have used equation \chidef\ to obtain the second equality.
Hence, using equations \alphaform\ we see that
$$
\det\chi^i{}_j={64\pi^2 G^2\kappa^2\over e^4}W^4\ ,
\eqn\sdeteqn
$$
and so equation \rhocpt\ is satisfied.

To summarise, we have demonstrated that the Einstein and field
equations can be reduced to a set of first--order Bogomol'nyi
equations given by
$$
\eqalign{
{dP\over d\rho}&=
\pm{e^4\over 2\kappa^2\sqrt{-g}}R^2\left(
R^2-\eta^2\right)g_{\theta\theta}-{e^2\over\kappa\sqrt{-g}}PR^2
g_{t\theta}
\cr
{dR\over d\rho}&=
{e^2\over 2\kappa\sqrt{-g}}R\left(
R^2-\eta^2\right)g_{t\theta}\mp{1\over\sqrt{-g}}PR
\cr
{1\over\sqrt{-g}}{dg_{t\theta}\over d\rho}&=
-{2e^2\pi G\over\kappa}\left( R^2-\eta^2\right)^2
\cr
{1\over\sqrt{-g}}{dg_{\theta\theta}\over d\rho}&=
\mp 8\pi G\left[
P\left( R^2-\eta^2\right) +\left( n+2s\right)\eta^2\right] -2\ .}
\eqn\fbogo
$$
Solutions to these Bogomol'nyi equations have spin and mass given by
$$
J=-{\pi\kappa\left( n+2s\right)^2\over e^2}\ ,
\qquad
M=\mp\pi\eta^2\left( n+2s\right)\ ,
\eqn\jmval
$$
where we choose the upper (lower) sign for $\left( n+2s\right)$
negative (positive) in order to obtain a positive value for the mass,
$M$. We believe that solutions to these Bogomol'nyi equations exist and
furthermore, guided by the flat space case [\flatt , \cugl], we
imagine that these solutions depend upon a single constant parameter
that will be determined by requiring the correct behaviour of the
fields and metric components as $\rho\rightarrow\infty$.

Note that, as claimed in the introduction, in contrast to the Abelian
case [\me] neither the angular momentum nor the mass are topologically
invariant. For each topological class (labelled by $n$) one can obtain
different values of $J$ and $M$ corresponding to different choices of
$s$. This is a manifestation of the fact that the topological charge
of \sun\ vortices is defined modulo $N$, whereas physical quantities
may depend upon the actual value of the magnetic flux associated with
this topological charge [\lozano]. This should not be overly
surprising since although two solutions belonging to the same homotopy
class are gauge equivalent at infinity, the gauge transformations
connecting solutions with different $s$ (but the same $n$) cannot be
well defined over all of the spacetime. Hence it is to be expected
that $J$ and $M$ differ according to the value of $s$. The most stable
vortex solutions are given by $n=1$, $s=0$ and $n=1$, $s=-1$.
Furthermore, due to the quantisation of the Chern--Simons coupling
constant, $\kappa$ given by equation \kappaquan\ we can express $J$ in
the form
$$
J={m\left( n+2s\right)^2\over 4}\ ,
\eqn\jquan
$$
where both $m$ and $s$ are integers. Thus, the angular momentum is
quantised. Note that this result (for $s=0$ and $s=-1$) agrees with the
value for the angular momentum in flat space given in [\flatt , \cugl].

\chapter{Conclusions}

In this article we have shown that, by considering an {\it ansatz} for
a cylindrically symmetric vortex solution, {\it both} the field
equations and the Einstein equations can be reduced to a set of four,
first--order Bogomol'nyi equations. In the flat space limit $\left(
G=0,\ g_{t\theta}=0\right)$ these Bogomol'nyi equations reduce to the
flat space Bogomol'nyi equations given in [\cugl]. In order to obtain
this set of equations it was necessary to choose an eighth--order,
non--renormalisable potential. Furthermore, we have demonstrated that
both the mass and the angular momentum of the vortex solutions are
{\it not} topologically invariant. It would be interesting to
numerically search for solutions to the Bogomol'nyi equations \fbogo .
Hopefully, this would justify the claims concerning the solutions made
above. Another open question is whether or not these vortex solutions
are stable. One way of investigating this problem is to embed this
theory in a supergravity theory. This would yield a set of
supersymmetry transformations that could be used to define a
supercovariant derivative. One could then perform a Witten--like
positive energy proof to obtain a Bogomol'nyi bound on the energy. We
imagine that solutions to the Bogomol'nyi equations above would
saturate this bound and hence be stable. In principle this appears to
be possible, but the construction of an appropriate supergravity
theory and the resulting supersymmetry transformations is not
immediately evident. This will be discussed in a future work.

\chapter{Acknowledgements}

The authors would like to thank B. Linet and G. Cl\' ement for
many useful and informative discussions that have facilitated this
article. This work was supported by a Royal Society European Science
Exchange Fellowship (L.A.J.L.) and by a grant from the Conselho
Nacional de Desenvolvimento Cient\'\i fico e Tecnol\'ogico (CNPq/Brazil)
(M.E.X.G.).

\refout

\end